\documentclass[journal = ancac3,manuscript=letter,layout=twocolumn]{achemso}

\usepackage{amssymb}
\usepackage[version=3]{mhchem} % Formula subscripts using \ce{}
\usepackage{gensymb}

\let\oldmaketitle\maketitle
\let\maketitle\relax
\author{Giorgio De Simoni}
\email{giorgio.desimoni@sns.it}
\affiliation[NEST]{NEST Istituto Nanoscienze-CNR  and Scuola Normale Superiore, I-56127 Pisa, Italy}
\author{Federico Paolucci}
\affiliation[NEST]{NEST Istituto Nanoscienze-CNR  and Scuola Normale Superiore, I-56127 Pisa, Italy}
\alsoaffiliation[INFN]
{INFN Sezione di Pisa, Largo Bruno Pontecorvo 3, 56127 Pisa, Italy}
\author{Claudio Puglia}
\affiliation[NEST]{NEST Istituto Nanoscienze-CNR  and Scuola Normale Superiore, I-56127 Pisa, Italy}
\alsoaffiliation[UNIPI]
{Dipartimento di Fisica dell'Universit\`a di Pisa - Largo Pontecorvo 3, I-56127 Pisa, Italy}
\author{Francesco Giazotto}
\affiliation[NEST]{NEST Istituto Nanoscienze-CNR  and Scuola Normale Superiore, I-56127 Pisa, Italy}
\email{francesco.giazotto@sns.it}
\title[]
    {Josephson Field-Effect Transistors based on All-Metallic Al/Cu/Al Proximity Nanojunctions}
\begin{document}

%%%%%%%%%%%%%%%%%%%%%%%%%%%%%%%%%%%%%%%%%%%%%%%%%%%%%%%%%%%%%%%%%%%%%
%% The "tocentry" environment can be used to create an entry for the
%% graphical table of contents. It is given here as some journals
%% require that it is printed as part of the abstract page. It will
%% be automatically moved as appropriate.
%%%%%%%%%%%%%%%%%%%%%%%%%%%%%%%%%%%%%%%%%%%%%%%%%%%%%%%%%%%%%%%%%%%%%
%\begin{tocentry}

%Some journals require a graphical entry for the Table of Contents.
%This should be laid out ``print ready'' so that the sizing of the
%text is correct.

%Inside the \texttt{tocentry} environment, the font used is Helvetica
%8\,pt, as required by \emph{Journal of the American Chemical
%Society}.

%The surrounding frame is 9\,cm by 3.5\,cm, which is the maximum
%permitted for  \emph{Journal of the American Chemical Society}
%graphical table of content entries. The box will not resize if the
%content is too big: instead it will overflow the edge of the box.

%This box and the associated title will always be printed on a
%separate page at the end of the document.
%\includegraphics[height= 3.5cm ]{TOC.pdf}

%\end{tocentry}

%%%%%%%%%%%%%%%%%%%%%%%%%%%%%%%%%%%%%%%%%%%%%%%%%%%%%%%%%%%%%%%%%%%%%
%% The abstract environment will automatically gobble the contents
%% if an abstract is not used by the target journal.
%%%%%%%%%%%%%%%%%%%%%%%%%%%%%%%%%%%%%%%%%%%%%%%%%%%%%%%%%%%%%%%%%%%%%
\twocolumn[
  \begin{@twocolumnfalse}
    \oldmaketitle
\begin{abstract}
We demonstrate proximity-based \textit{all-metallic} mesoscopic superconductor-normal metal-superconductor (SNS) field-effect controlled Josephson transistors (SNS-FETs) and show their full characterization from the critical temperature $T_c$ down to 50 mK in the presence of both electric and magnetic field. The ability of a static electric field -applied by mean of a lateral gate electrode- to suppress the critical current $I_s$ in a proximity-induced superconductor is proven for both positive and negative gate voltage values. $I_s$ reached typically about one third of its initial value, saturating at high gate voltages. The transconductance of our SNS-FETs  obtains values as high as 100 nA/V at 100 mK. On the fundamental physics side, our results suggest that the mechanism at the basis of the observed phenomenon is quite general and does not rely on the existence of a true pairing potential, but rather the presence of superconducting correlations is enough for the effect to occur. On the technological side, our findings widen the family of materials available for the implementation of all-metallic field-effect  transistors to \textit{synthetic} proximity-induced superconductors.

\end{abstract}
 \end{@twocolumnfalse}
]

%%%%%%%%%%%%%%%%%%%%%%%%%%%%%%%%%%%%%%%%%%%%%%%%%%%%%%%%%%%%%%%%%%%%%
%% Start the main part of the manuscript here.
%%%%%%%%%%%%%%%%%%%%%%%%%%%%%%%%%%%%%%%%%%%%%%%%%%%%%%%%%%%%%%%%%%%%%
%\textbf{Keywords}: superconducting proximity effect, electric field effect, Josephson effect, supercurrent transistor

\section{}
Superconducting electronics, \textit{i.e.} based on electronic circuits made of superconducting materials, is nowadays a well-established industrial platform to implement fast and energy efficient information architectures. It offers quantum and classical computation devices \cite{Nakamura1999,Yamamoto2003,Wendin2017,Larsen2015,DeLange2015,Casparis2016,DeSimoni2018NanoLetters} and quantum tools whose field of application includes current limiters, electronic filters, routers for communication networks, analogue-to-digital converters operating at GHz frequencies \cite{Mukhanov2004,Mukhanov2011}, magnetometers, digital receivers and photon detectors. These devices have no competitors in the semiconductor world in terms of signal-to-noise ratio, energy efficiency, speed and frequency of operation. Also, superconducting  wires can carry sub-nanosecond current pulses without distortions and with low cross-talk for distances easily overcoming the millimeter\cite{Likharev1991a}.
\begin{figure}[h!]
\includegraphics[width= \columnwidth]{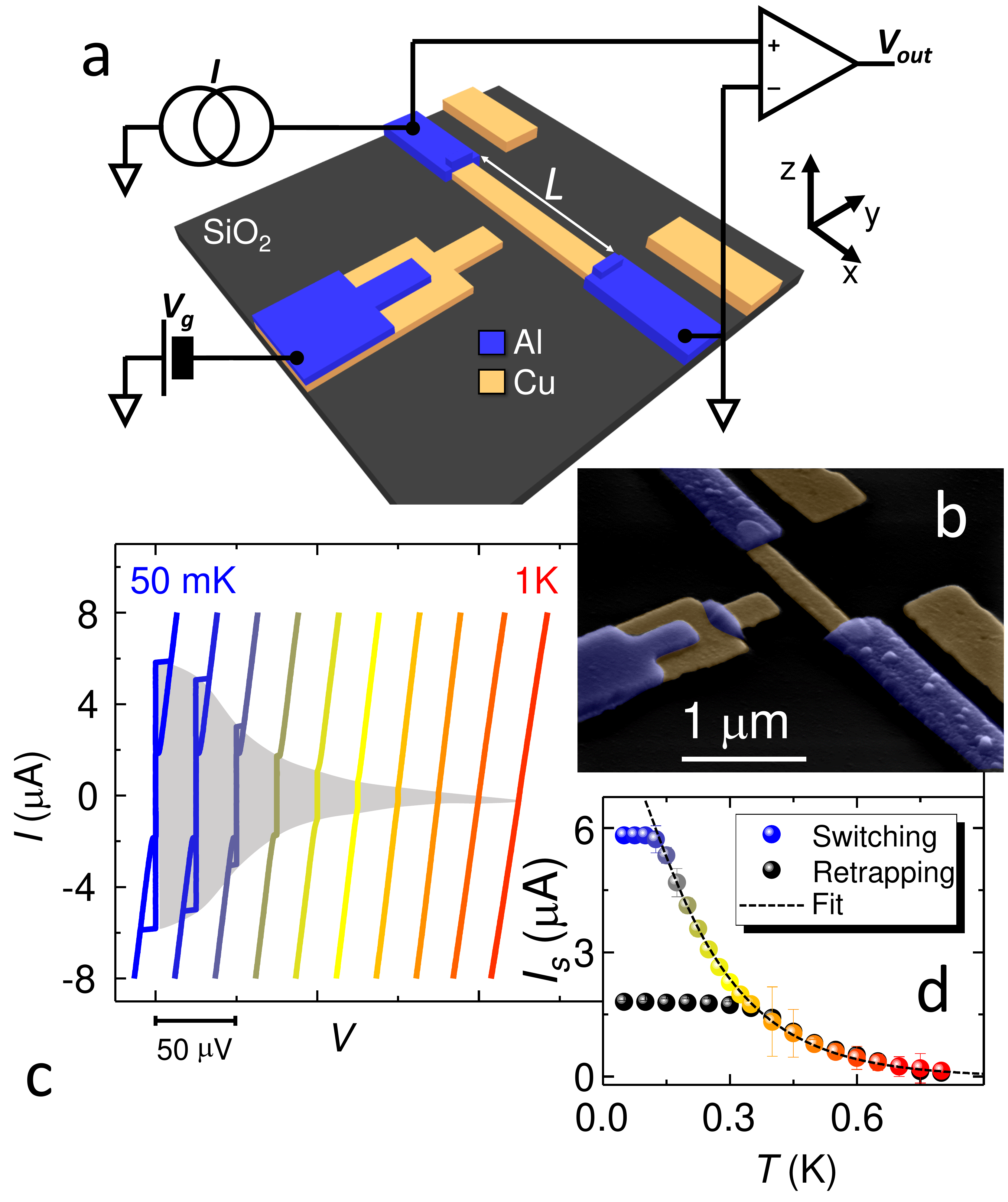}
\caption{\label{fig1} \textbf{a:} 3-dimensional scheme of an all-metallic mesoscopic SNS-FET. $L$ denotes the length of the junction. In blue it is shown the superconducting material (Al); in orange the normal metal (Cu). A scheme of the electrical measurements setup is also shown. \textbf{b:} False color tilted scanning-electron-microscope image of a B-type sample. The color scheme is the same of panel a. Replicas of the electrodes due to angle evaporation deposition technique are also visible. \textbf{c:} Selected current vs. voltage characteristics of an A-type SNS-FET with $L$=800 nm at several bath temperatures from 50 mK to 950 mK, with a step of 100 mK. Curves are horizontally offset for clarity. Clear hysteretic behaviour in the switching current is present up to $\sim$350 mK. The junction normal-state resistance extracted from the $I(V)$ curves is $R_N\simeq$4.4 $\Omega$. \textbf{d:} Switching and retrapping current as a function of temperature for an A-type SNS-FET. The color scheme of the switching current is the same of panel c. A fit of the switching current is also shown (dashed line). Fit yielded a Thouless energy $E_{Th}\simeq$6.6$\mu$eV and an ideality factor $\alpha$=0.63 with a coefficient of determination $R^2$=0.9994 (see text).}
\end{figure}

The conventional \textit{knobs} to operate superconducting devices are provided by the Josephson effect and the quantization of the magnetic flux: information is usually stored through the presence or absence of a single flux quantum (SFQ) in a superconducting loop accounting for logical state 1 or 0. Such states, on the one hand, can be written by means of magnetic fields applied through external magnets or  \textit{on-chip} coils. On the other hand a Josephson junction (JJ), \textit{i.e.} a weak link between two superconductors, can be used as non linear circuital element to be switched from the superconducting to the resistive state by increasing the circulating current above its  critical current $I_s$, that is the maximum dissipationless current sustained by the JJ. The transition to the resistive state yields a change in the magnetic flux threading a superconducting loop and therefore allows to perform a digital logic operation.

A possible interface circuitry between superconducting and complementary metal-oxide-semiconductor  (CMOS) electronics is provided nowadays by non-Josephson superconducting devices like, \textit{e.g.}, the n-Trons \cite{McCaughan2014,Zhao2017}: three terminal  circuits consisting of a superconducting strip in which superconductivity can be quenched by an extra-current-pulse injected by a third galvanically-connected terminal. Differently from JJs, n-Trons can be conveniently designed to exhibit the desired impedance and output voltage signal in the normal state \cite{Zhao2017}.

On this background, a separate discussion must be reserved to conventional electric-field-effect based superconducting electronics: almost every device owning to this class, essentially, relies on charge-carrier concentration control through the application of a gate voltage $V_g$ to a superconducting-proximitized semiconductor \cite{Clark1980,Doh2005,Xiang2006,Paajaste2015} or to a high-critical-temperature \cite{Fiory1990} superconducting conduction channel. This results in a modulation of the superconducting critical temperature $T_c$ \cite{Fiory1990,Nishino1989}, of the normal-state resistance $R_N$ and of $I_s$ \cite{Okamoto1992,Mannhart1993,Mannhart1993_B,Takayanagi1985,Akazaki1996,Jespersen2009,Abay2014}. Low charge-density is a strict requirement for these effects to occur, in order to prevent electrostatic screening to cancel the field at wire surface with no significant effects on its conduction properties. 

Surprisingly, although Bardeen-Cooper-Schrieffer (BCS) and Fermi-liquid theories \cite{Larkin1963,Virtanen2019} forbid field-effect to be functional on superconducting metals due to their high surface charge density, it was recently demonstrated that the electric field couples with the Cooper condensate in BCS all-metallic transistors allowing, on the one hand, a control (down to its full suppression) of the supercurrent \cite{DeSimoni2018,Paolucci2018,Paolucci2019} and, on the other hand, affecting the superconducting phase-drop across a gated Josephson junction (JJ) \cite{Paolucci2019_SQUID}. Moreover, under the action of the electrostatic field, all-metallic supercurrent transistors exhibit a behavior which cannot be assimilated to conventional field-effect superconducting devices as the device charge density and normal-state resistance remains totally unchanged. On the technological side, these discoveries promise the realization of all-metallic field-effect superconducting devices such as, \textit{e.g.}, metallic gatemons \cite{Larsen2015,DeLange2015,Casparis2016} or the so-called EF-Trons, \textit{i.e.}, field-effect n-Trons\cite{Paolucci2019}, that could take advantage, differently from the high-$T_c$ superconductors or their semiconducting counterparts, of a monolithic structure requiring a simple single-step fabrication process based on abundant and easy-to-manage materials like aluminum or titanium. On the fundamental physics side, this phenomenology, which still does not have a microscopic description, indicate an obscure zone in our understanding of the nature of the BCS state, and demand for a profound theoretical and experimental re-investigation of superconductivity. Among the many open questions, here we try to experimentally tackle the main point whether the field effect requires, to be effective on supercurrent, a genuine superconducting material or if, instead, the effect manifests itself also in proximitized metallic superconductor\cite{Likharev1991,Dubos2001,Courtois2008,Wang2009,Garcia2009,Ronzani2013}s: we report on the demonstration of mesoscopic superconductor-normal metal-superconductor field-effect controlled Josephson transistors (SNS-FETs). We show their full characterization from $T_c$ down to 50 mK in the presence of both electric and magnetic fields. Our results show the ability of an externally-applied static electric field to largely tune $I_s$ in a proximity-induced superconductor similarly to what previously observed in genuine superconductors \cite{DeSimoni2018,Paolucci2018,Paolucci2019}, therefore,  strongly suggesting that the mechanism at the basis of the observed phenomenon does not rely on the presence of a true pairing potential but that, instead, the existence of superconducting correlations seems to be enough for the effect to occur. These experimental findings expand the family of metals available for the realization of metallic superconducting field-effect transistors, and are of great relevance with regards to the physics of Josephson coupling in SNS proximity junctions, which is one of the most robust manifestation of coherence in mesoscopic devices\cite{Likharev1991}.

\begin{figure}[t!]
\includegraphics[width= \columnwidth]{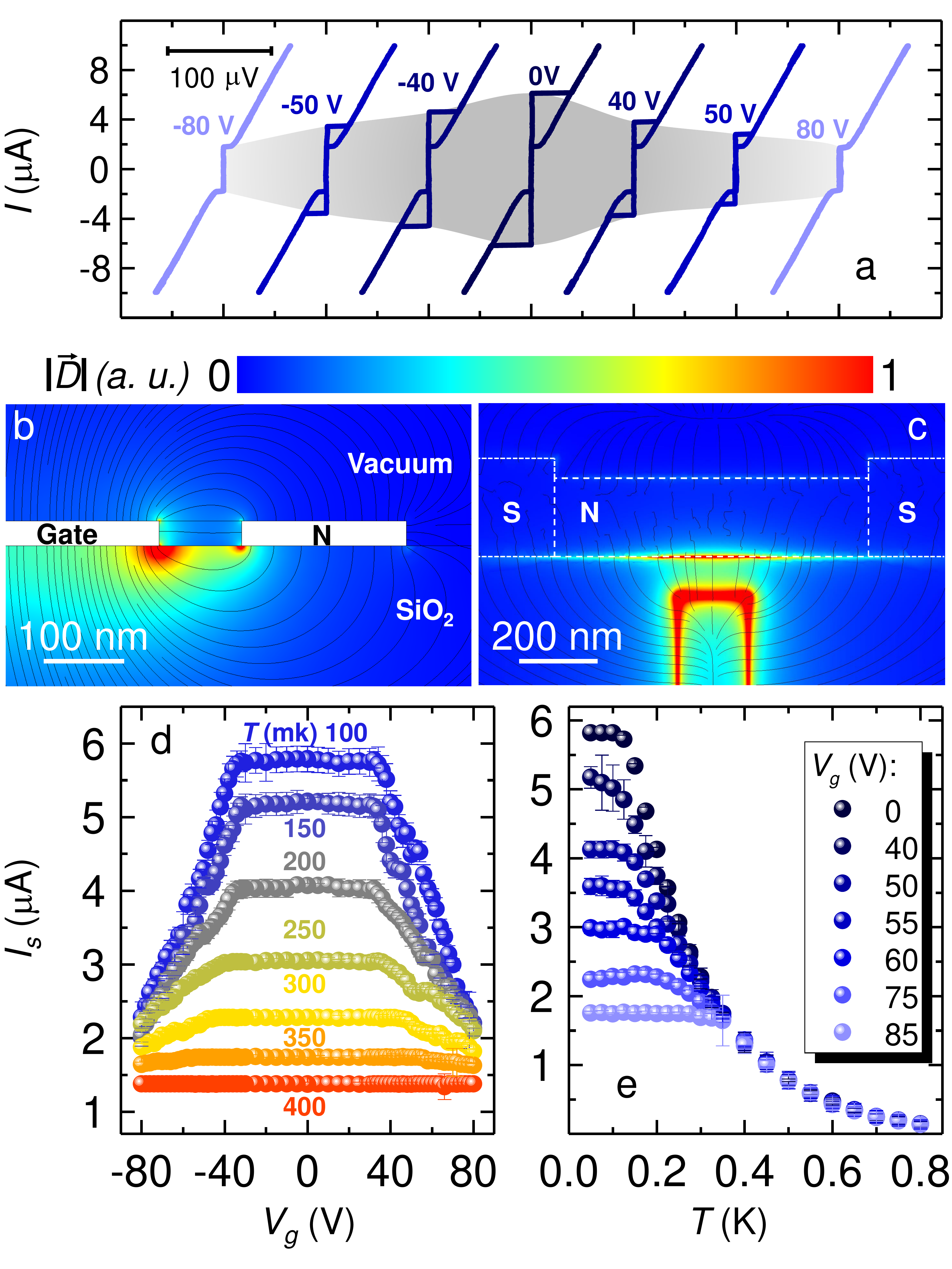}
\caption{\label{fig2} \textbf{a:} Selected current vs. voltage characteristics of an A-type SNS-FET with $L$=800 nm at gate voltages $V_g$ from -80 V to 80 V. Curves are horizontally offset for clarity. \textbf{b,c:} Contour plot of the result of FEM calculation of the modulus of the displacement field $| \vec D|$ for a device with geometry of A-type SNS-FET. $| \vec D|$ was calculated by solving the equation $\vec \nabla \vec D=\sigma$ when a voltage $V_g$=1 V was applied at the gate electrode. The S and N sections of the JJ were modeled as perfect conducting domains. $\vec D$ is plotted in arbitrary units ranging from 0 (blue) to 1 (red). Streamline of the field are also shown. Panel b refers to $yz$ located at the center of the weak link. Panel c refers to the $xy$ plane 10 nm below the substrate surface. Axis definition is reported in Fig. \ref{fig1}a. \textbf{d:} Average $I_s(V_g)$ extracted from 50 repetitions of $I(V)$ measurements at constant gate voltage $V_g$ of a representative A-type SNS-FET for several bath temperatures $T$. \textbf{e:} Average $I_s(T)$ extracted from 50 repetitions of $I(V)$  measurements at constant $T$ of a representative A-type SNS-FET for several $V_g$ values.}
\end{figure}

\section{Results and Discussion}
Our SNS-FETs consist of Al/Cu/Al planar gated junctions. The Cu normal metal wire was 200 nm wide and 30 nm thick. Several devices were realized with inter-S-electrode spacing $L$ of the Cu weak link equal to 0.8 $\mu$m (A-type sample), 1.0 $\mu$m (B-type sample) and 1.2 $\mu$m (C-type sample). The 180-nm-wide Cu gate electrode was separated by a distance of about 100 nm from the normal-metal wire. Further details of nano-fabrication process and of the measurement technique is reported in the \textit{Methods/Experimental} section. A  3-dimensional representation of a typical SNS-FET comprising a scheme of the setup used for the 4-wire electrical characterization is depicted in Fig. \ref{fig1}a, while a tilted false color scanning electron-microscope image of a B-type device is shown in Fig. \ref{fig1}b.

Figure \ref{fig1}c shows the current-voltage [$I(V)$] characteristics of a representative A-type SNS-FET at several temperatures from 50 mK up to 950 mK. The curves are horizontally offset for clarity. For temperatures smaller than 750 mK, the $I(V)$s exhibit the Josephson effect with a switching current $I_s$ of 5.8 $\mu$A at 50 mK  and a normal-state resistance $R_N\simeq4.4$ $\Omega$. Stemming from electron heating in the N region once the junction switches to the normal state\cite{Garcia2009,Ronzani2013,Dubos2001,Courtois2008,Uchihashi2011}, a clear thermal hysterical behavior is present when the $I(V)$ is collected forward and backward with a retrapping current $I_r \simeq 1.86$ $\mu$A. A plot of $I_s$ and $I_r$ \textit{versus} bath temperature $T$ is shown in Fig. \ref{fig1}d. As usually observed in SNS JJs\cite{Garcia2009,Dubos2001}, the difference between $I_s$ and $I_r$ decreases as $T$ is increased and vanishes at $T\simeq$350 mK. 

The Thouless energy ($E_{Th}$) of the weak link, which is the characteristic energy scale for electrons diffusing through a finite-sized conductor and in proximity-induced superconductivity, was determined from $I_s(T)$ by a least-square minimization fitting procedure with the following  relation which holds for $k_B T\gtrsim 5E_{Th}$\cite{ZaikinA.;Zharkov1981,Dubos2001,Courtois2008}:
%\times\\
\begin{equation}
%\begin{multline}
    I_s(T)=\gamma
    \sum_{n=0}^\infty \frac{T \sqrt{\frac{2\omega_n}{E_{Th}}}\Delta^2(T)e^{-\sqrt{\frac{2\omega_n}{E_{Th}}}}}{[\omega_n+\Omega_n+\sqrt{2(\Omega_n^2+\omega_n\Omega_n)}]^2},
    \label{eq1}
%\end{multline}
\end{equation}
with $\omega_n(T)=(2n+1)\pi k_B T$ and $\Omega_n(T)=\sqrt{\Delta^2(T)+\omega_n^2(T)}$. $\Delta$ is the BCS aluminum pairing potential, while $\gamma$ is a factor defined as $\alpha$64$\pi$ ${k_B T}$/$q {R_N}$ where $k_B$, $q$ and $\alpha$ are the Boltzmann constant, the electron charge and a suppression coefficient accounting for non ideal transmissivity through the S-N interface, respectively. The fit yielded $E_{Th}$=6.6$\mu$eV and $\alpha$=0.63. Goodness of the fit was evaluated computing the coefficient of determination $R^2=1-\frac{\sum_{i}(y_i-\bar{y})^2}{\sum_{i}(y_i-f_i)^2}$=0.9994, where $y_i$ and $f_i$ are respectively data and fitted points and $\bar{y}$ is the mean of data points. The resulting fit curve is superimposed on top of the experimental data in Fig. \ref{fig1}d (dashed line). Our weak links operate in the diffusive regime and within the long-junction limit, holding for $E_{Th}\ll\Delta\simeq180\mu$ eV and posses an effective length $L_{eff}=\sqrt{\hbar D/E_{Th}}\sim$900 nm, which is deduced through the diffusion coefficient of Cu $D$=1/($\rho$ $q^2 v_F$)$\simeq 0.008$ m$^2$/s where $v_F$=1.56$\times$10$^{47}$ J$^{-1}$m$^{-3}$  is the density of states at the Fermi energy and $\rho$ the  resistivity of the Cu wire. 

\begin{figure}[t!]
\includegraphics[width= \columnwidth]{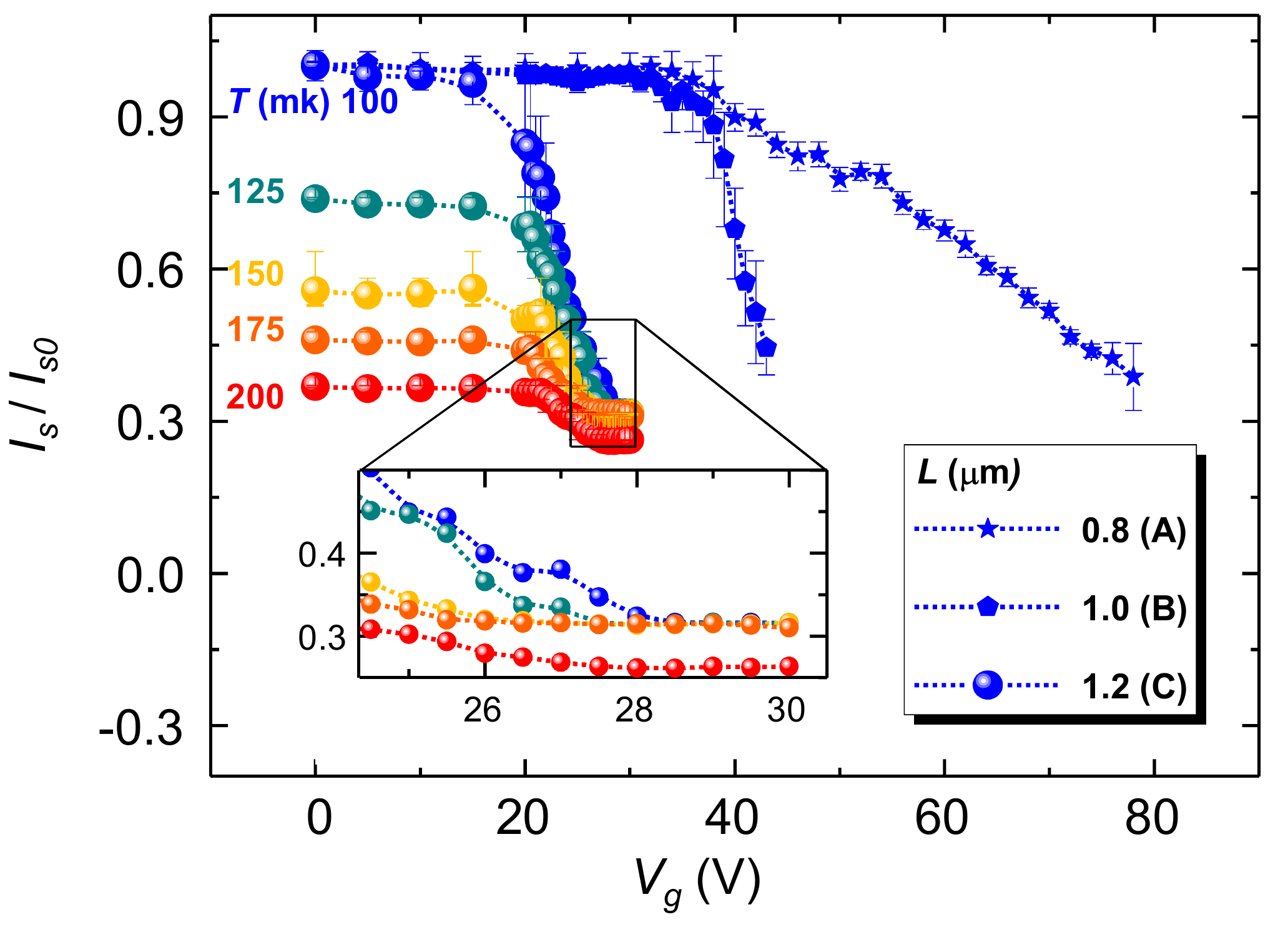}
\caption{\label{fig3} Average normalized $I_s(V_g)/I_{s0}$ extracted from 50 repetitions of $I(V)$ measurements at constant gate voltage $V_g$ of representative A-type (same device of Fig. \ref{fig1} and \ref{fig2},  stars), B-type (pentagons, $I_{s0}=5.0$ $\mu$A) and C-type (spheres, $I_{s0}=1.4$ $\mu$A) SNS-FETs. For C-type device thermal characterization is also shown. \textbf{Inset:} a blow-up of the saturation region $I_s(V_g)$ at high $V_g$.}
\end{figure}

A direct way to prove the ability of the electrostatic field to tune the Josephson coupling of the SNS-FET is the acquisition of $I(V)$ curves for several values of gate voltage $V_g$ and different bath temperatures. Figure \ref{fig2}a shows the current-voltage [$I(V)$] characteristics of a representative A-type SNS-FET collected at 100 mK for several $V_g$ values ranging from -80 V to 80 V. The curves are horizontally offset for clarity. For $|V_g |\gtrsim$40 V a clear suppression of the switching current $I_s$ was observed, while the normal-state resistance $R_N$ and $I_r$ were completely unaffected by the gate voltage. When $I_s(V_g)$ became lower than $I_r$ the hysteretic behaviour was not present anymore (Fig. \ref{fig2}a).

Considering that the superconducting-like properties of the Cu wire were inherited from the Al banks, a preliminary 3-dimensional finite element method (FEM) calculation of the intensity of the electric displacement vector $\vec D$ was performed by solving the equation  $\vec \nabla \vec D=\sigma$, describing the Gauss’s law in the differential form, for the geometry of an A-type SNS-FET ($\sigma$ is the charge density), in order to asses that the predominant effect on $I_s$ originated from a direct action of the electric field on the N wire. In our calculation the gate electrode and the S and N sections of the device were approximated as ideal conductor boundaries with the constrain of $V=V_g=1 V$ on the gate surfaces and $V=0$ on the JJ surfaces. Figures \ref{fig2}b and c report the contour plots of $|\vec D |$ calculated respectively in the $yz$ plane located at the center of N wire, and in the $xy$ plane lying 10 nm below the substrate surface (see definition of axes in Fig. \ref{fig1}a). As expected, due to the higher dielectric constant of the SiO$_2$ layer, $|\vec D |$ is more intense into the substrate than into vacuum. Notably, $| \vec D |$ decays pretty fast along the weak link, being suppressed by an order of magnitude at the S-N interfaces. This fact allows to safely attribute any reduction of the device switching current to a direct action of the gate on the N-section of the junction and to exclude field-effect-driven weakening of superconductivity in the Al leads as the main mechanism at the bases of the observed phenomenology.

\begin{figure}[t!]
\includegraphics[width= \columnwidth]{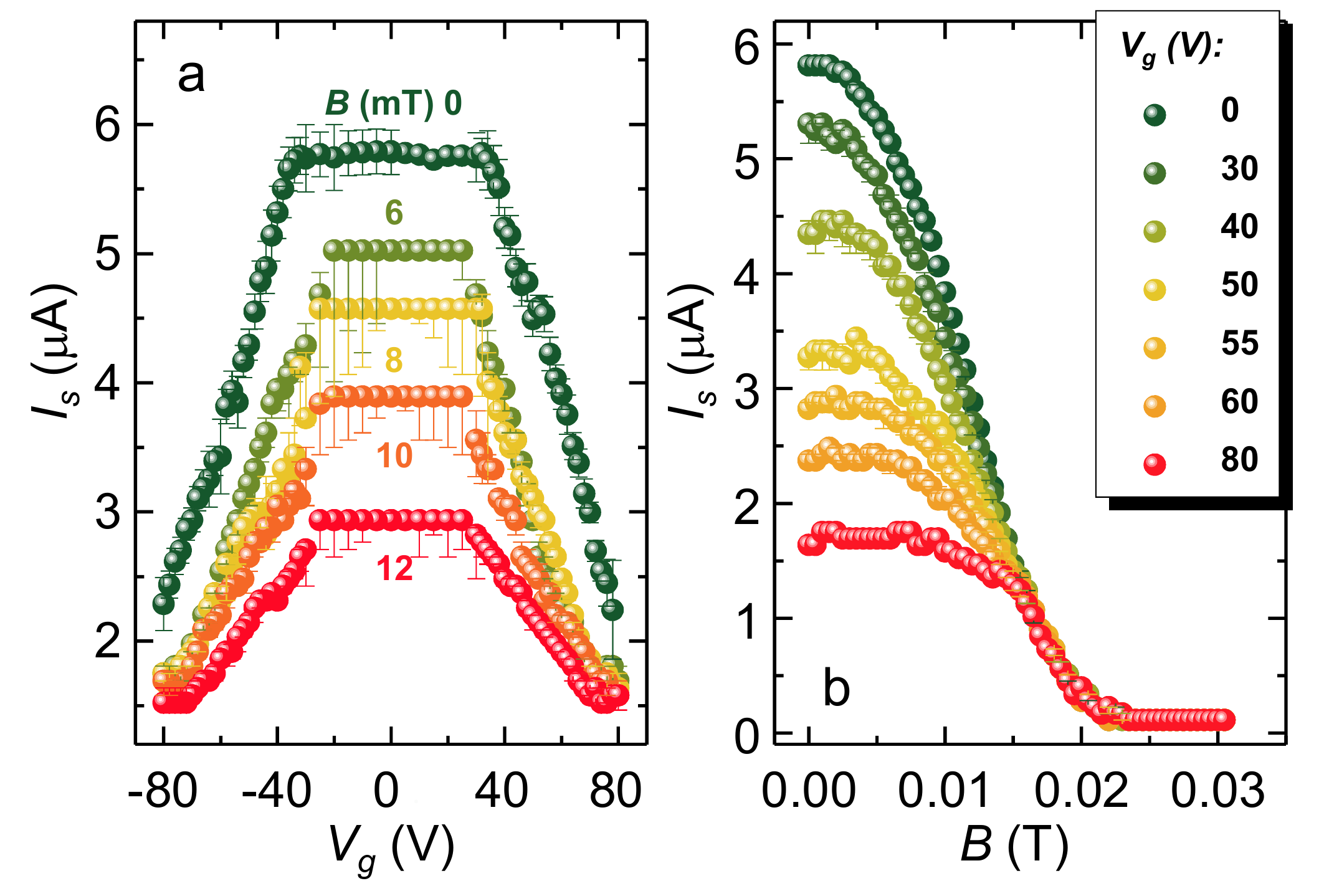}
\caption{\label{fig4} \textbf{a:} Average $I_s(V_g)$ extracted from 50 repetitions of $I(V)$ measurements at constant gate voltage $V_g$ of a representative A-type SNS-FET for several values of the magnetic field $B$. \textbf{b:} Average $I_s(T)$ extracted from 50 repetitions of $I(V)$  measurements at constant $B$ of a representative A-type SNS-FET for several values of $V_g$. The measurements in panel a and b  were performed at $T=100$ mK}
\end{figure}

The $I_s$ values extracted from the $I(V)$ performed on the A-type SNS-FET are shown in Fig. \ref{fig2}d. At fixed bath temperature, $I_s(V_g)$ monotonically decreases without reaching full suppression in the explored voltage range, differently from what previously observed on genuine superconducting Dayem bridge devices\cite{Paolucci2018,Paolucci2019}. Also, in contrast to high-Tc superconductors and proximitized-semiconductor field effect transistors, $I_s$ suppression is almost symmetric with respect of the sign of $V_g$ thereby indicating a bipolar behavior in the electric field, and $R_N$ is totally unaffected by $V_g$. The bipolarity excludes any charge depletion/accumulation process as main driving mechanism for $I_s$ reduction. By increasing the values of the temperature yields a lower value of $I_{s0}=I_s(V_g=0)$ and a larger range of ineffectiveness of the electric field on $I_s$, \textit{i.e.} the plateau of constant $I_s $ widens. This latter behavior resembles the results obtained on Ti and Al superconducting FETs\cite{DeSimoni2018,Paolucci2018,Paolucci2019}. These findings seem to indicate a direct link between the electric field applied to the Cu wire and the observed $I_s$ suppression, with apparently no obvious relation with the electric field experienced by the superconducting Al leads. This observation suggests that  the presence of superconducting correlation is enough for the manifestation of electrostatic field control of the supercurrent.

For $T\simeq 400$ mK, that is around one third of aluminum critical temperature, field-effect becomes completely ineffective. This behaviour can be better appreciated by looking at the $I_s(T)$  curves measured for several $V_g$ values (see Fig. \ref{fig2}e): while for $T \gtrsim 350$ mK all the curves are overlapped, at lower temperature $I_s$ significantly deviates from the unperturbed case (\textit{i.e.}, $V_g=0$) showing an $I_s$ plateau which widens when $V_g$ is increased. Although no microscopic model exists for field-effect controlled SNS devices, our measurements seem to suggest that, as the Cu wire is pushed toward its normal state -by either increased bath temperature or through the application of the electric field- it partially recovers a metallic behaviour where no field-effect can be observed. This is consistent with the hypothesis of dissipative puddles arising in the weak-link as a consequence of the externally applied electric field\cite{Paolucci2019}.

The above consideration founds a qualitative confirmation in the comparison of the $I_s(V_g)$ characteristics of SNS-FETs with different lengths. Figure \ref{fig3} reports the comparison of the normalized critical current $I_s(V_g)/I_{s0}$ for representative A-type, B-type and C-type devices. On one side, the plateau of $I_s$ turns out to decrease as $L$ increases. In addition, in the shown C-type device, a very clear saturation region for high $V_g$ values was observed. We emphasize that this feature is strictly peculiar of SNS junctions and has no counterparts in genuine superconducting Dayem bridges in which full suppression of $I_s$ was observed\cite{DeSimoni2018,Paolucci2018,Paolucci2019}. These results suggest that, in weaker proximity-superconductors, the impact of the electric field is initially more relevant, but as the system approaches the resistive state a threshold is reached above which no further reduction of $I_s$ can be observed. This facts points toward the existence of a non-trivial relation between the length of the junction and its resilience to the electric field or, in other words, between the junction Thouless energy and the ability of the field effect to affect the supercurrent. Yet, the presence of a saturated $I_s$ at high $V_g$ seems to exclude the possibility that $I_s$ suppression stems from a direct hot-electron injection into the weak link due to a leakage current between the gate electrode and the JJ\cite{Roddaro2011,Morpurgo1998,Savin2004}.

The interplay between electric and magnetic fields $B$ on $I_s$ in SNS-FETs was investigated by acquiring $I_s$ in the presence of a constant magnetic field applied along z-axis (see Fig. \ref{fig4}a). In contrast to measurements at different temperatures, no clear widening of $I_s$ plateau was observed, but rather a slight non-monotonic narrowing is present. This feature is strictly peculiar of SNS-FET, since it was not reported in genuine superconductor metallic FETs \cite{DeSimoni2018,Paolucci2018,Paolucci2019}. For $B\geq 8$ mT, a clear saturation of $I_s$ at high gate voltage was observed in the SNS-FET. $I_s(B)$ measurements at constant $V_g$ are shown in Fig. \ref{fig4}b: similarly to the analogous thermal characterization, for $B\gtrsim 15$ mT the curves at different gate voltages overlaps. These observations confirm a damping of the electrostatic effect on the weak link as it approaches the normal state.  

\begin{figure}[t!]
\includegraphics[width= \columnwidth]{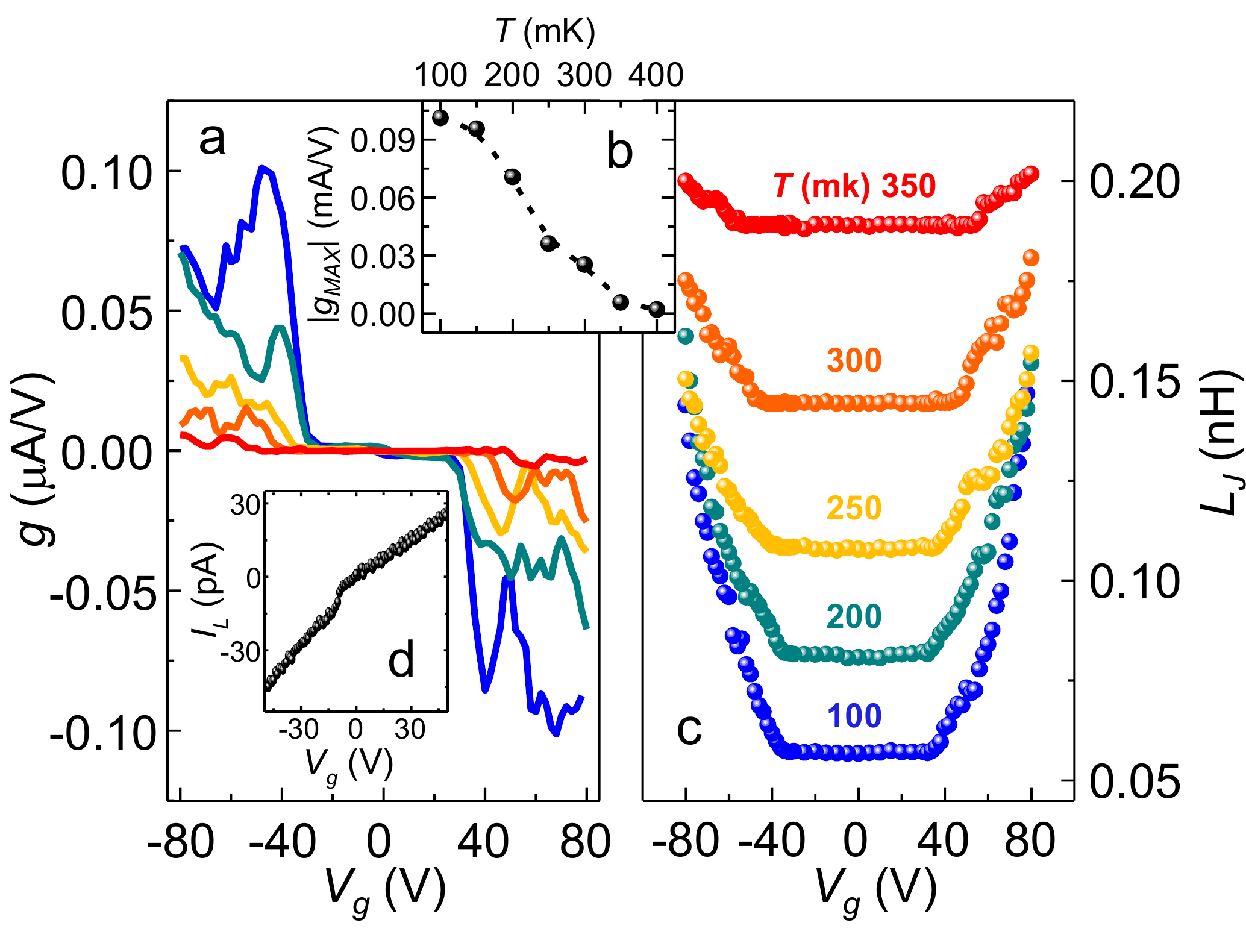}
\caption{\label{fig5} \textbf{a:} Transconductance $g$ as a function of gate voltage $V_g$ for different values of the temperature $T$ measured on a representative A-type device. \textbf{b}: Absolute value of the maximum of transcoductance $|g_{MAX}|$ as a function of temperature for the same device of panel a. \textbf{c:} Josephson inductance $L_J$ as a function of $V_g$ for different values of $T$ of the same device of panel a \textbf{d:} Leakage current $I_L$ measured in our setup for an A-type SNS-FET. A similar behaviour was observed also in B-type and C-type devices.}
\end{figure}

We finally comment on the SNS-FET performance in terms of the gate-JJ impedance, the transconductance and the Josephson inductance. To exclude the presence of a direct hot-electron injection into the weak link, the leakage current $I_L$  between the gate and the drain electrode was measured. $I_L$ was found to be always at most of the order of few tens of pA (see Fig. \ref{fig5}d), with a typical measured gate-JJ impedance in our setup of a few T$\Omega$, also compatible with a slow charging-discharging of low-pass filters and parasitic capacitance in the measurement setup. The plot of the transconductance $g(V_g)=dI_s/dV_g$ acquired for an A-type device is reported for several  temperatures in Fig. \ref{fig5}a, and provides the conventional figure of merit relevant for technological applications. Due to the bipolar behavior of our SNS-FETs, the transconductance is an odd function of $V_g$. In addition, $g$ has a strong temperature dependence which reflects in the evolution of the absolute value of its maximum $g_{MAX}$ \textit{versus} $T$, as shown in Fig. \ref{fig5}b. In stark contrast with the genuine superconducting case\cite{Paolucci2018,Paolucci2019}, $g_{MAX}$ has no constant range in temperature, and monotonically decreases vanishing at 400 mK. The gate-dependent suppression of $I_s$ results into an increase of the Josephson inductance, $L_J = \hbar/(2qI_s)$ (see Fig. \ref{fig5}c). Typical $L_J$ excursion ranges from a few tens to a few hundreds of pH. We wish to stress that, although the performances of the SNS-FETs here reported are almost one order of magnitude worse than those obtained on Dayem bridge field-effect devices, such figures of merit could be improved up to a large extent by improving the design of the device by merely decreasing the distance between the gate electrode and the weak-link and by exploring the effect of the deposition and encapsulation of the transistor in high-dielectric-constant insulators such as, \textit{e. g.}, SrTiO$_3$ \cite{Lippmaa1999}. We would like also to highlight that, by the same engineering activity, it will be possible to match the gate impedance in order to allow for high-speed commutation of the our transistor, whose frequency of operation is expected to be limited essentially by the parasitic capacitance of the gate electrode and by the typical time-scale of the superconducting to normal-state transition. Finally, we wish to point out  that the operation temperature of our SNS-FETs can be easily raised from the sub-Kelvin temperatures up to a few Kelvin by choosing the material of the S leads among higher-Tc superconductors such as \textit{e. g.} vanadium, niobium, lead, or niobium nitride. These considerations let us to foresee a potential applicability of our mesoscopic SNS-FETs for the realization of all-metallic EF-Trons.

\section{Conclusions}
In summary, we have demonstrated \textit{all-metallic} mesoscopic SNS Josephson field-effect transistors. Our results show the ability of the electrostatic field to tune the Josephson coupling even in the presence of a metallic proximitized superconductor, thereby suggesting that no true pairing potential is needed for the effect to occur. In contrast to the Dayem bridge geometry so far explored, SNS JJs seem not to allow a gate-driven full suppression of the supercurrent. This fact, which is still lacking a theoretical interpretation, seems to indicate a role of the normal-metal nature of the weak link in determining the resilience of the Josephson coupling to the electric field. From the technological point of view, our findings potentially increase the number of metals suitable for the realization of Josephson transistors, they offer the possibility to exploit the dependence of $I_s$ suppression on the Thouless energy as a further knob to tune the response of the device to the electric field, and let to envisage performances already on par with high-Tc and proximitized-semiconductor-based superconducting transistors.

\section{Experimental Methods}
Our SNS-FETs consist of Al/Cu/Al planar gated junctions fabricated by a single step electron beam lithography of a suspended resist mask [a double layer consisting of methyl methacrylate (MMA) and Poly methyl methacrylate (PMMA)] and angle resolved evaporation of metals\cite{Cord2006} onto an oxidized silicon wafer (the SiO2 is 300 nm thick) in an ultra high vacuum electron-beam evaporator with base pressure of about 10$^{-11}$ torr. The 200-nm-wide normal metal wire, consisting of 5-nm-thick Ti adhesion layer (evaporated at a rate 1.2 Å/s) and 30-nm-thick Cu film (evaporated at 1.5 Å/s), was evaporated without tilting the sample. The 100-nm-thick superconducting Al banks were evaporated by tilting the substrate at 30\degree. The 180-nm-wide  gate electrode was separated by a distance of about 100 nm from the normal-metal Cu wire, with which it shared the material composition, since it was deposited in the same evaporation step.

The electrical characterization of our mesoscopic transistors was performed by four-wire technique in a  filtered $^3$He-$^4$He dilution fridge by setting a low-noise current bias and measuring the voltage drop across the weak links with a room temperature pre-amplifier.

%%%%%%%%%%%%%%%%%%%%%%%%%%%%%%%%%%%%%%%%%%%%%%%%%%%%%%%%%%%%%%%%%%%%%
%% The "Acknowledgement" section can be given in all manuscript
%% classes.  This should be given within the "acknowledgement"
%% environment, which will make the correct section or running title.
%%%%%%%%%%%%%%%%%%%%%%%%%%%%%%%%%%%%%%%%%%%%%%%%%%%%%%%%%%%%%%%%%%%%%
%\begin{acknowledgement}
\section{Acknowledgement}
We acknowledge F. S. Bergeret, A. Braggio, M. Cuoco, V. Golovach, J. D. Sau, P. Solinas, E. Strambini and P. Virtanen for fruitful discussions. The authors acknowledge the European Research Council under the European Union's Seventh Framework Programme (COMANCHE; European Research Council Grant No. 615187) and Horizon 2020 and innovation programme under grant agreement No.  800923-SUPERTED. The work of G.D.S. and F.P. was partially funded by the Tuscany Region under the FARFAS 2014 project SCIADRO. The work of F.P. was partially supported by the Tuscany Government (Grant No. POR FSE 2014-2020) through the INFN-RT2 172800 project. 

%\end{acknowledgement}

%%%%%%%%%%%%%%%%%%%%%%%%%%%%%%%%%%%%%%%%%%%%%%%%%%%%%%%%%%%%%%%%%%%%%
%% The same is true for Supporting Information, which should use the
%% suppinfo environment.
%%%%%%%%%%%%%%%%%%%%%%%%%%%%%%%%%%%%%%%%%%%%%%%%%%%%%%%%%%%%%%%%%%%%%
%\begin{suppinfo}

%This will usually read something like: ``Experimental procedures and
%characterization data for all new compounds. The class will
%automatically add a sentence pointing to the information on-line:

%\end{suppinfo}

%%%%%%%%%%%%%%%%%%%%%%%%%%%%%%%%%%%%%%%%%%%%%%%%%%%%%%%%%%%%%%%%%%%%%
%% The appropriate \bibliography command should be placed here.
%% Notice that the class file automatically sets \bibliographystyle
%% and also names the section correctly.
%%%%%%%%%%%%%%%%%%%%%%%%%%%%%%%%%%%%%%%%%%%%%%%%%%%%%%%%%%%%%%%%%%%%%
\section{Author Contributions}
G.D.S., F.P, and C.P. fabricated the samples. G.D.S. and C.P. performed the experiment and analyzed the data with inputs from F.G. F.G. conceived the experiment. G.D.S. wrote the manuscript with input from all the authors. All of the authors discussed the results and their implications equally.

%\section{Associated Content}
%A pre-print version of this work is available\cite{DeSimoni2019_SNSarXiv} .

%\bibliography{SNSFET}

\providecommand{\latin}[1]{#1}
\makeatletter
\providecommand{\doi}
  {\begingroup\let\do\@makeother\dospecials
  \catcode`\{=1 \catcode`\}=2 \doi@aux}
\providecommand{\doi@aux}[1]{\endgroup\texttt{#1}}
\makeatother
\providecommand*\mcitethebibliography{\thebibliography}
\csname @ifundefined\endcsname{endmcitethebibliography}
  {\let\endmcitethebibliography\endthebibliography}{}
%\begin{mcitethebibliography}{46}

\end{document}